\documentclass[aps,prb,showpacs]{revtex4}
\usepackage{amsbsy,latexsym}
\usepackage{amsfonts}
\usepackage{amssymb}
\usepackage[mathscr]{eucal}

\begin{document}

\title{Scaling laws for the 2d 8-state Potts model with Fixed Boundary Conditions}
\author{M.~Baig}
\affiliation{Grup de F{\'\i}sica Te{\`o}rica \& IFAE, Facultat de
Ci{\`e}ncies, Edifici Cn, Universitat Aut{\`o}noma de Barcelona,
08193 Bellaterra (Barcelona) Spain}
\author{R.~Villanova }
\affiliation{Matem{\`a}tica Aplicada, DEE, Universitat Pompeu Fabra,
08028 Barcelona, Spain}

\begin{abstract}
We study the effects of frozen boundaries in a Monte Carlo
simulation near a first order phase transition. Recent
theoretical analysis of the dynamics of first order phase
transitions has enabled to state the scaling laws governing the
critical regime of the transition. We  check these new scaling
laws performing a Monte Carlo simulation of the 2d, 8-state spin
Potts model. In particular, our results support a pseudo-critical
$\beta(L)$ finite-size scaling of the form
$\beta(\infty)+a_1/L+a_2/L^2$, instead of
$\beta(\infty)+\theta_1/L^d+\theta_2/L^{2d}$. Moreover, we obtain a latent
heat, $\Lambda_{\rm FBC}=0.294(11)$, which does not coincide with the latent
heat analytically derived for the same model if periodic boundary conditions are
assumed, $\Lambda_{\rm PBC}=0.486358\ldots$
\end{abstract}

\maketitle

\section{Introduction}

The introduction of computer simulation methods has been a
breakthrough in the study of phase transitions in lattice models.
The rapid increase on the computers power has enabled to analyze
with great accuracy the scaling laws, governed by the critical
exponents, and even the corrections to these scaling laws. In such
analysis, finite size effects must be taken into account
carefully~\cite{Binder}. One of these effects, the disturbance
from the boundary, has usually been dismissed by the adoption of
periodic boundary conditions (PBC). Nevertheless, in some
situations the adoption of periodic lattices may not be adequate,
either for practical or theoretical reasons. This is the case of
the free boundary conditions used in the analysis of free
surfaces, or the so called boundary fields, used in the analysis
of wetting phenomena. In the present paper we will focus on a
particular election of the boundary conditions, the so called
fixed boundary conditions (FBC), which have been recently applied
to spin models~\cite{BEJ,BEJ2} and gauge models~\cite{BF}.

Second order phase transitions exhibit universality. For this
reason, all the details of the system near the phase transition
point become irrelevant for the critical exponents. By contrast,
first order phase transitions are not universal and, hence, all
details of a simulation must be considered carefully. This
includes, in particular, the choice of boundary conditions. What
are the appropriate set of scaling laws for a first order phase
transition with FBC? This is the question we address in this
paper. Starting from the theoretical analysis of Borgs and
Kotecky~\cite{BK,BK2} and the diploma Thesis of Medved~\cite{M} on
the dynamics of first order transitions, we present the finite
size scaling laws applicable to the case of FBC and we check them
performing a numerical simulation of the 2-d 8-state spin Potts
model~\cite{Potts} with FBC.

The paper is divided as follows. In section II a brief summary of
some recent simulations where FBC have been adopted serves as a
motivation for a detailed analysis of the finite size scaling
laws which are also presented and discussed. Section III is
devoted to a discussion of our numerical simulation and the
results that we have obtained. In section IV we analyze our
results in the light of the scaling laws presented in Sec.~II.
Finally, in section V we give some concluding remarks.

\section{Fixed Boundary Conditions}

\subsection{The motivation for FBC}

Recently, the so called {\em gonihedric spin models}have been
proposed as a laboratory to study discrete versions of string
theories~\cite{SW}. All these simulations have been performed
imposing standard periodic boundary conditions on a
three-dimensional lattice. Nevertheless, in some cases three
intersecting inner planes of spins were fixed to break the large
energy degeneracy of the hamiltonian\cite{BEJ,BEJ2}. Due to the
periodicity of the boundaries, this is equivalent to fix the
spins belonging to the six planes of the 2-d boundary of the 3-d
cube formed by the spins. Since for a certain range of the
coupling parameter, in particular for $\kappa=0$, the transition
is clearly of first order, the analysis of the finite size
effects should have been done using the scaling laws presented in
this paper. We expect that the application of the FBC scaling laws
may overcome some anomalies recently observed in the analysis of
this transition~\cite{BV}.

Another situation where the knowledge of the FBC scaling laws
seems to be crucial is the issue of the triviality of lattice QED.
Indeed, it has been claimed that the formation of artificial
monopole structures, which close over the boundaries, in a
simulation of the 4-d U(1) gauge model may be responsible for
turning the phase transition of this model from second to first
order~\cite{LJ}. To avoid this problem, originated probably by an
incorrect choice of the boundaries, it was suggested to perform
the Monte Carlo simulations on a lattice with a topology of an
sphere. Along these lines, Baig and Fort\cite{BF} proposed the
adoption of FBC to simulate an spherical topology. Effectively,
to fix all the variables belonging to the 3-d border to unity is
the higher dimensional equivalent of converting a 2-d plane
square lattice to the 2-d surface of a sphere by collapsing the
lines of the border to a single point. Nevertheless, to
discriminate between a first or a second order nature for a
transition, an accurate analysis of data produced is necessary,
and, in particular, this will be only possible if one knows for
certain the applicable scaling laws.

\subsection{The scaling laws}

Although speaking properly no critical exponents can be defined
for first order phase transitions, it is usual to define a set of
characteristic exponents, together with a set of scaling laws
borrowed from those of the second order phase transitions.
The pioneering work of Privman
and Fisher~\cite{Privman}, Binder and Landau~\cite{Binder2} and
Challa {\em et al.}~\cite{Challa}, provided a phenomenological understanding
of the scaling for first order transitions.
A more rigorous theoretical justification for these first order scaling laws was
presented by Borgs and Kotecky~\cite{BK,BK1}. The formulation of
its applicability to finite size scaling expressions in terms of
the lattice size was the work of Borgs, Kotecky and
Miracle-Sol\'{e}~\cite{BKS} and, independently, of Janke~\cite{Janke}.
But in all these developments the existence of periodic
boundary conditions was assumed. Recently, though,  Borgs and
Kotecky~\cite{BK2} have extended their analysis to include
surface effects in addition to the standard volume effects which
govern first order transitions. Following this work,
Mendev~\cite{M} has deduced the scaling laws for the spin Potts
model in the presence of surface effects, in particular adopting
boundary conditions other than the periodic ones.

Following the general analysis of Mendev~\cite{M}, finite size
scaling laws in terms of the lattice size for the case of fixed
boundary conditions can easily be deduced. They are summarized in
Table~\ref{tab:scal-laws}, together with the standard laws for
periodic conditions. In the rest of this paper we will check
these modified scaling laws with the results of our numerical
simulation.

It should be noticed that the suggestion that in the case of free
boundary conditions every transition is shifted by a $1/L$
correction term caused by surface effects is quite old.
Binder~\cite{Binder3}, for instance,
reports on a series of experimental results~\cite{Experiments}
supporting this conclusion.

%
%

\begin{table}
\centering \caption[{\em Nothing.}] {{\em Scaling laws for
Periodic and Fixed Boundary Conditions.}} \vspace{3ex}
\begin{tabular}{|l|c|c|} \hline
    & P.B.C. & F.B.C. \\ \hline
  $\beta_{c}^{peaks}(L) =$ & $\beta_{c}(\infty)+ \frac{\theta_1}{L^{d}}+O(\frac{1}{L^{2d}})$ &
   $\beta_{c}(\infty)+ \frac{a_1}{L}+O(\frac{1}{L^{2}})$ \\
 $C_{max}(L) =$ & $\gamma_{0}+\gamma_{2}L^{d}+O(\frac{1}{L^{d}})$ &
   $c_{0}+c_{2}L^{d}+O(L^{d-1})$ \\
$\chi_{max}(L)= $ & $\delta_{0}+\delta_{2}L^{d}+O(\frac{1}{L^{d}})$ &
   $e_{0}+e_{2}L^{d}+O(L^{d-1})$ \\
$B_{min}(L) = $ & $\Phi_{0}+\frac{\Phi_{1}}{L^{d}}+O(\frac{1}{L^{2d}})$
& $B_{0}+\frac{B_{1}}{L}+O(\frac{1}{L^{2}})$\\ \hline
\end{tabular}
\label{tab:scal-laws}
\end{table}

\section{Numerical simulation}

To test the scaling laws of Table~\ref{tab:scal-laws}, we have performed
a numerical simulation of the 2-d 8-state spin Potts model defined by the
partition function
\begin{equation}
Z_{potts}=\sum_{\{\sigma_{i}\}}e^{-\beta E},
\end{equation}
where the energy is
\begin{equation}
E=-\sum_{\langle ij\rangle}\delta_{\sigma_{i}\sigma_{j}},
\; \;\;\;\; (\sigma_{i}=1,...,8),
\end{equation}
with $\beta=J/kT$ in natural units. It is well known that this
model exhibits a first order phase transition \cite{Wu} and
for this reason it has been chosen as a test model in several
previous studies.

Fixed boundary conditions have been implemented along the lines
stated by Baig and Fort~\cite{BF}. In a 2-d grid with points
labeled by $(n_x,n_y)$, all spins corresponding to the lattice
points $(1,n_y)$ and $(n_x,1)$ for $n_x,n_y = 1,...,L$ have been
fixed during all the simulation at its initial values $\sigma=1$.
With this precaution, the structure of the program, that
implements PBC, assures the persistence of the frozen boundary.

We have performed the lattice updating applying a well tested head
bath algorithm. During the simulation we recorded time series files
for the energy $E$ and the magnetization $M$ defined as
\begin{equation}
         M=\frac{q \; \mbox{max}\{n_i\}-L^d}{q-1},
\end{equation}
where $q=8$ and $n_i$ is the number of spins in a given
orientation.

Table~\ref{tab:statistics} summarizes the details of the
simulations that have performed from $L=70$ up to $L=350$. The
number of production Monte Carlo sweeps varies from $n_{\rm
prod}=6~000~000$ for $L=70$, to  $n_{\rm prod}=32~700~000$ for
$L=350$. Since we took measurements only every $n_{\rm flip}=8$
sweeps,  the number of total measurements per run is $n_{\rm
meas}=n_{\rm prod}/n_{\rm flip}$. We left at least $20 \, n_{\rm
flip} \tau_{\rm e}$ thermalization sweeps before taking
measurements\cite{Sokal,Sokal2,Janke2}. To estimate the
autocorrelation time of energy measurements $\tau_{\rm e}$, we
have applied two different methods. First, we use the fact that
$\tau_{\rm e}$ enters the error estimate $\epsilon_{\rm
JK}=\sqrt{2\; \tau_{\rm e}/n_{\rm meas}} \, \epsilon_{\rm naive}$
for the mean energy $<E>$ of $n_{\rm meas}$ correlated energy
measurements of variance
$$
 \epsilon_{\rm naive}^2=\sum_{j=1}^{n_{\rm meas}}(<E>-E_j)^2/(n_{\rm meas}-1).
$$
\noindent The ``true" error estimate $\epsilon_{\rm JK}$ is
obtained splitting the energy time-series into 50 bins, which
were in their turn jackknived~\cite{jack,jack2} to decrease the
bias in the analysis. The second way of obtaining $\tau_{\rm e}$
is by a direct computation of the integrated autocorrelation time
$$
 \tau_e^{\rm int} = \frac{1}{2} + \sum_{j=1}^{k_{\rm max}-1}
 (1-j/k_{\rm max}) \frac{\frac{1}{k_{\rm max}-(j-1)} \;
 \sum_{i=1}^{k_{\rm max}-(j-1)} (E_i-<E>)(E_{i+j}-<E>)}
 {<E^2>-<E>^2},
$$
where  $k_{\rm max}$ is a suitable cut-off~\cite{Sokal} around $6
\, \tau_e^{\rm int} < k_{\rm max} < 10 \, \tau_e^{\rm int}$. The
corresponding error in $\tau_e^{\rm int}$ is derived from the
{\em a priori} formula $\sqrt{2\, (2\, k_{\rm max} +1)/n_{\rm
meas}} \, \tau_e^{\rm int}$.

%
%
\begin{table}[htbp]
\centering \caption[{\em Nothing.}]
 {{\em  Monte Carlo parameters of the simulation. $L^2$ is the
        lattice size, $n_{\rm therm}$ the number of Monte Carlo
        sweeps during thermalization, and $n_{\rm prod}$ the number
        of production runs. Measurements were taken every $n_{\rm flip}=8$
        Monte Carlo sweeps for all the simulations.}}
\vspace{3ex}
\begin{tabular}{r|lrrrcrrr}
\multicolumn{1}{c}{$L$}  & \multicolumn{1}{c}{$\beta_{MC}$}  &
\multicolumn{1}{c}{$n_{\rm therm}$}     &
\multicolumn{1}{c}{$n_{\rm prod}$}    &
\multicolumn{1}{c}{$\tau_{\bf \rm e}$} &
\multicolumn{1}{c}{$\tau_{\bf \rm e}^{\rm int}$} &
\multicolumn{1}{c}{$\frac{n_{\rm therm}/n_{\rm flip}}{\tau_{\bf
\rm e}}$} &
\multicolumn{1}{c}{$\frac{n_{\rm prod}/n_{\rm flip}}{2 \, \tau_{\bf \rm e}}$} \\ \\

\hline
   70  & 1.3343    & 100 000  &  6 000 000 &    144 & 128(12)& 87 & 2604 \\ \\
   84  & 1.3363    & 100 000  &  6 000 000 &    208 & 240(25)& 60 & 1803 \\ \\
  100  & 1.3378    & 150 000  &  8 000 000 &    357 & 394(38) & 53 & 1441 \\ \\
  126  & 1.33909   & 250 000  &  8 000 000 &    883 & 847(122) & 35 &  567 \\ \\
  150  & 1.3398    & 400 000  &  10 000 000 &  1320 & 1341(215) & 38 &  474 \\ \\
  200  & 1.3407    & 900 000  &  12 000 000 &  4664 & 5434(1582) & 24 &  161 \\ \\
  226  & 1.34102  & 1 200 000 &  16 000 000 &  7287 & 6991(1476) & 21 &  138 \\ \\
  250  & 1.341205  & 1 600 000 & 18 000 000 &  9072 & 9700(2454)& 22 &  124 \\ \\
  278  & 1.34138   & 2 200 000 & 18 800 000 & 11743 & 16969(5058)& 23 &  100  \\ \\
  300  & 1.34146   & 3 000 000 & 22 000 000 & 15429 & 27765(12969)& 24 &   89 \\ \\
  350  & 1.34162   & 4 000 000 & 32 700 000 & 25632 & 53623(29055)& 20 &   80 \\ \\
\end{tabular}
\label{tab:statistics}
\end{table}

\begin{figure}[p]
 \vspace*{18cm}
   \includegraphics{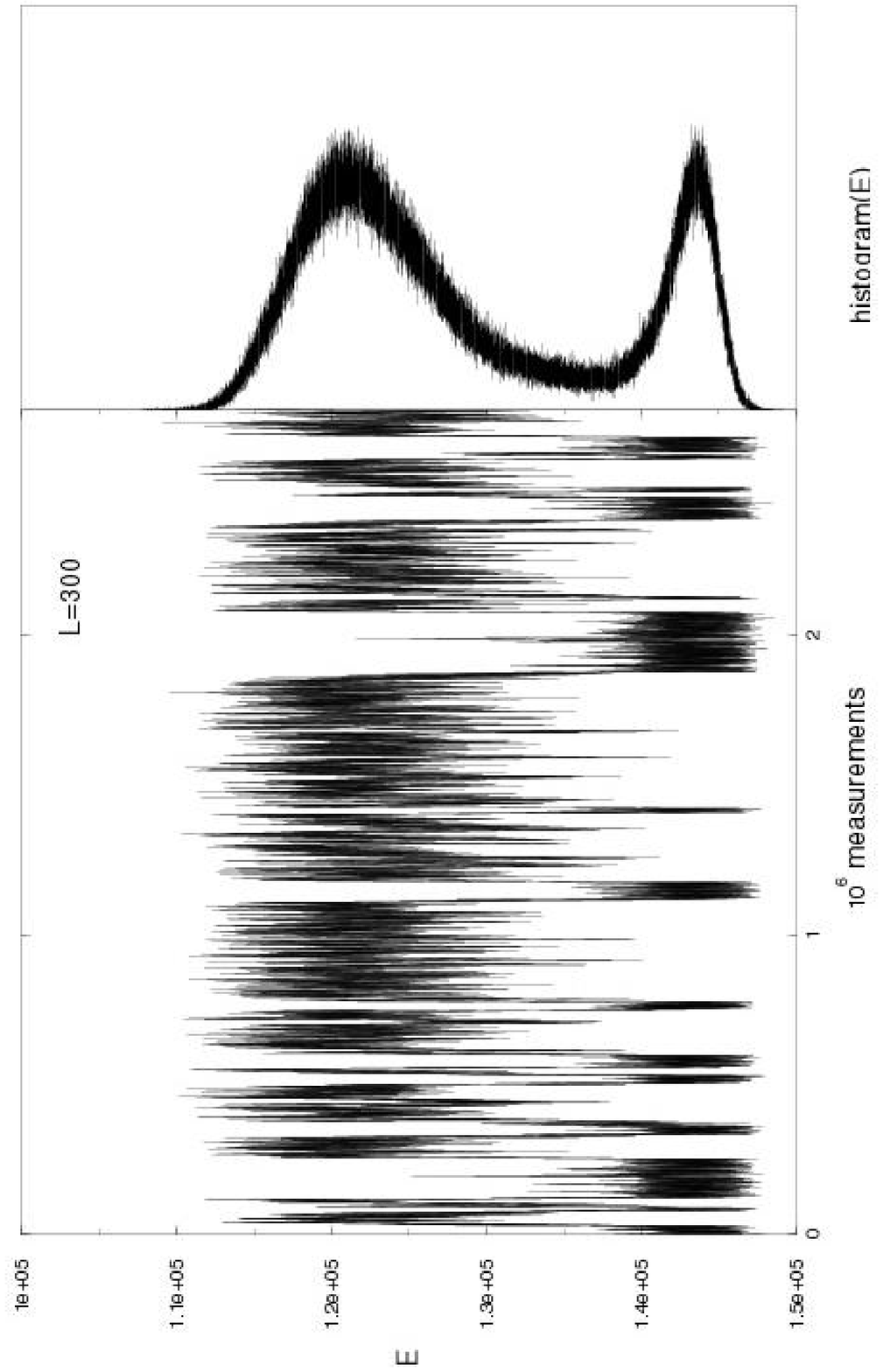}
 \caption{Energy time series for $L=300$ and $\beta_{MC}=1.34146$}
 \label{fig:time-series}
\end{figure}

In Fig.~\ref{fig:time-series} we present the energy time-series
for the $L=300$ and $\beta_{MC}=1.34146$ simulation run. The
expected characteristic behaviour for a first order phase
transition can be clearly seen. The system remains in one of the
two coexisting phases for a long period of time. The energy
histogram for the full series is also presented in this figure.
The similar height of the two peaks confirms that the simulation
was performed very near the pseudo-critical inverse temperature.

\begin{figure}[ht]
 \vspace*{9cm}
  \includegraphics{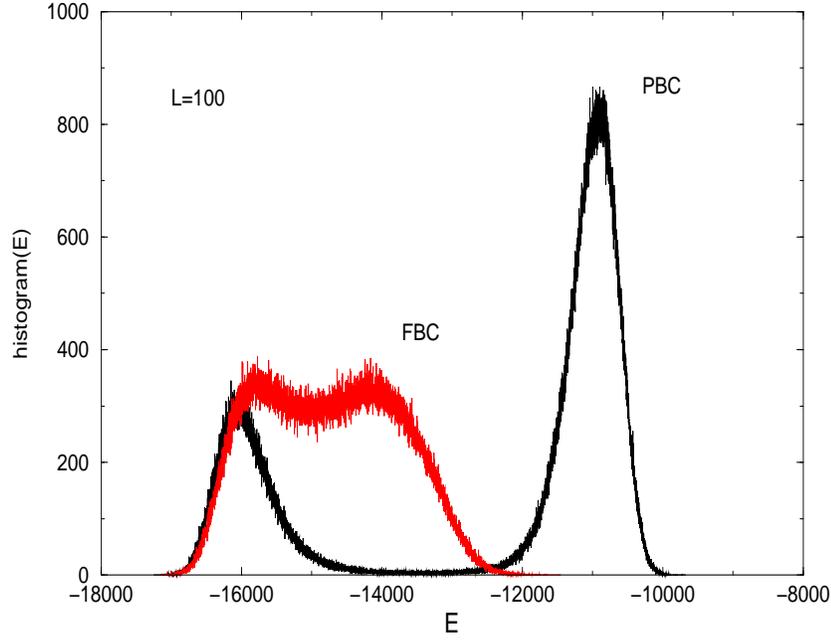}
 \caption{Energy histograms for $L=100$ and $\beta_{MC}^{PBC}=1.342027$
 and $\beta_{MC}^{FBC}=1.3378$.}
 \label{fig:hist2}
\end{figure}

It is instructive to compare the energy histograms corresponding
to the adoption of fixed or periodic boundary conditions. To this
end we have performed two different Monte Carlo runs, close to the
respective pseudo-critical inverse temperatures, which are
 $\beta_{MC}^{PBC}=1.342027$ and
$\beta_{MC}^{FBC}=1.3378$ for a lattice size $L=100$. These
simulations has been done using $8~000~000$ production sweeps,
with $n_{\rm flip}=8$, discarding the initial $250 000$ ($150
000$) sweeps in the case of PBC (FBC) for the thermalization of
the system. Both histograms can be seen in Fig.~\ref{fig:hist2}.
They show the characteristic two-peaks structure. Nevertheless,
the latent heat, i.e., the separation between the maximum of the
two peaks, is clearly smaller for Fixed Boundary Conditions. This
qualitative observation suggest that a simple analysis of the
energy histograms of a true first order phase transition
simulated with Fixed Boundary Conditions might be misleading.
Effectively, if the lattice size is not large enough the energy
histogram could show (apparently) a single peak and, in
consequence, one can get the erroneous conclusion that the model
exhibit a second order phase transition. Nevertheless, even with
FBC the evolution of the energy histograms when the size of the
system increases shown in Fig.~\ref{fig:hist2} ($L=100$), and in
Fig.\ref{fig:time-series} ($L=300$), exhibit the expected
behaviour of a first order transition. This observation may be
relevant in the interpretation of the analysis of Baig and
Fort~\cite{BF}, where a disappearance of a two peaks structure
was observed when FBC were imposed to the system~\cite{N1}.

In addition to the qualitative analysis of the histograms, we have
computed the specific heat, magnetic susceptibility and the Binder
kurtosis parameter at nearby values of $\beta_{MC}$ by means of
standard reweighting techniques~\cite{Swendsen}. They are defined as
\begin{eqnarray}
C(\beta) & = & \frac{\beta^{2}}{L^{2}}(\langle
E^{2}\rangle-\langle
E\rangle^2), \\
\chi(\beta) & = & \frac{\beta^{2}}{L^{2}}(\langle
M^{2}\rangle-\langle
M\rangle^2), \\
B(\beta) & = & 1-\frac{\langle E^4 \rangle}{3 \langle
E^2\rangle^2}.
\end{eqnarray}

In table~\ref{tab:extrema} we show the extrema of the magnitudes
above defined, together with their pseudo-critical inverse
temperatures. The error bars of these quantities have been
estimated splitting the time-series data into 50 bins, which were
jackknived to decrease the bias in the analysis of reweighted
data.
%
%
\begin{table}[htbp]
\centering \caption[{\em Nothing.}]
 {{\em Extrema for the (finite lattice) specific heat, $C_{\rm max}$,
      the susceptibility, $\chi_{\rm max}$, and the energetic Binder
      parameter, $B_{\rm min}$, together with their respective
      pseudo-critical inverse temperatures.}}
\vspace{3ex}
\begin{tabular}{r|lrlrll}
\multicolumn{1}{c}{$L$}  & \multicolumn{1}{c}{$\beta_{\rm
max}^{C}$} & \multicolumn{1}{c}{$C_{\rm max}$} &
\multicolumn{1}{c}{$\beta_{\rm max}^{\chi}$} &
\multicolumn{1}{c}{$\chi_{\rm max}$}    &
\multicolumn{1}{c}{$\beta_{\rm min}^{B}$} &
\multicolumn{1}{c}{$B_{\rm min}$}  \\
\hline
   70  & 1.334469(53) & 89.26(96) & 1.334212(52) & 95.3(1.2)
       & 1.333966(52) & 0.660221(74) \\ \\
   84  & 1.336360(46) & 124.6(1.4) & 1.336215(45) & 143.8(1.7)
       & 1.336040(46) & 0.660441(71) \\ \\
  100  & 1.337705(34) & 171.3(1.9) & 1.337620(33) & 210.5(2.5)
       & 1.337492(33) & 0.660657(69) \\ \\
  126  & 1.339124(33) & 268.4(4.1) & 1.339081(33) & 355.8(5.7)
       & 1.339000(33) & 0.660774(94)  \\ \\
  150  & 1.339905(25) & 391.5(7.4) & 1.339881(25) & 550(11)
       & 1.339821(25) & 0.66058(12)   \\ \\
  200  & 1.340747(23) & 757(16) & 1.340739(22) & 1144(25)
       & 1.340704(22) & 0.66006(15)   \\ \\
  226  & 1.341046(18) & 1006(27) & 1.341042(18) & 1554(43)
       & 1.341014(18) & 0.65981(19)   \\ \\
  250  & 1.341229(15) & 1285(29) & 1.341225(15) & 2033(47)
       & 1.341203(15) & 0.65950(17)  \\ \\
  278  & 1.341379(12) & 1695(42) & 1.341377(12) & 2735(69)
       & 1.341358(12) & 0.65900(20)  \\  \\
  300  & 1.341493(12) & 2083(51) & 1.341491(12) & 3413(86)
       & 1.341475(12) & 0.65856(21)  \\ \\
  350  & 1.3416496(92) & 3176(77) & 1.3416490(92) & 5338(130)
       & 1.3416373(92) & 0.65754(23) \\ \\
\end{tabular}
\label{tab:extrema}
\end{table}

\section{Scaling laws analysis}

Once we have the results from the numerical simulation on finite
lattices, we can proceed to analyze the data imposing the scaling laws
of Table~\ref{tab:scal-laws}.

\subsection{Analysis of the pseudo-critical inverse temperature}

\begin{figure}[p]
 \vspace*{9cm}
   \includegraphics{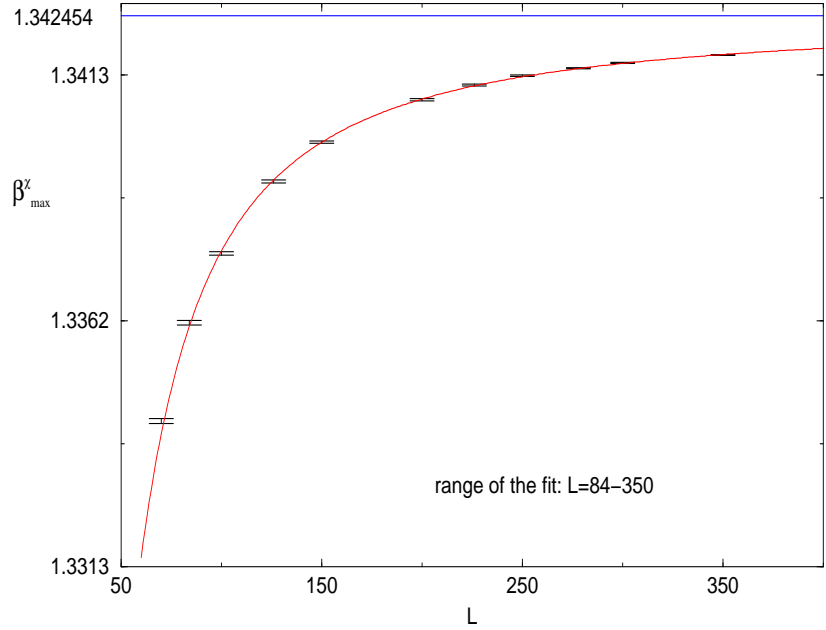}
 \caption{Finite-size scaling analysis of the pseudo-critical
  $\beta^{\chi}_{\rm max}$ in the range $L=84-350$ by means of the ansatz
  $\beta^{\chi}_{\rm max}(L)=\beta_c+\alpha_1/L+\alpha_2/L^2$. The infinite volume
  critical point obtained from the fit is $\beta_c=1.342478(38)$, with a
  goodness-of-fit $Q=0.13$.}
 \label{fig:beta-fit}
\end{figure}

In Table~\ref{tab:beta-fits} we present the results of fitting the
pseudo-critical betas of $C_{\rm max}$, $\chi_{\rm max}$ and
$B_{\rm min}$ to the ansatz $\beta_c+a_1/L+a_2/L^2$ suggested by
the finite-size scaling laws presented in
Table~\ref{tab:scal-laws}. Notice that we have performed two set
of fits, one for the full range $84\leq L\leq 350$, and a second
including only results from the lattice sizes $100 \leq L\leq
350$. Notice that the fits are extremely good even for the initial
range $84\leq L\leq 350$, but they improve slightly if $L=84$ is
discarded. Remember that reasonable fits should have a
goodness-of-fit~\cite{NR}, $Q$, above 0.05.
Fig.~\ref{fig:beta-fit} depicts the fit for $\beta_{\chi}^{\rm
max}(L)$ in the range $84\leq L\leq 350$. The {\em exact} critical
inverse temperature for the 2d 8-state Potts model is
$\beta_c(exact)=\ln(1+\sqrt(8))=1.342454 \ldots$. Our results of
Table~\ref{tab:beta-fits} are in perfect agreement with this
value.

We have also fitted our data to the ansatz
$\beta^{peaks}_c=\beta(\infty)+\theta_1/L^2+\theta_2/L^4$ corresponding to
the PBC finite-size scaling law. Even though the
goodness-of-fit, $Q$, obtained does not allow to discard the
fits, the infinite volume $\beta_c(\infty)$ resulting from them
do not coincide with the exactly known value, showing that this
ansatz is unsuitable. I.e. for the $\beta^C_{\rm max}(L)$ in the
range $84\leq L\leq 350$, the fit produces $Q=0.10$ and
$\beta_c(\infty)=1.342063(11)$ and for $\beta^{\chi}_{\rm max}(L)$
in the range $100 \leq L\leq 350$, the results are $Q=0.21$ and
$\beta_c(\infty)=1.342079(13)$.

%
%
\begin{table}[htbp]
\centering \caption[{\em Nothing.}]
 {{\em  Pseudo-critical inverse temperature fits. Q is the
        goodness-of-fit. Recall that the
        exact critical inverse temperature for the model is
        $\beta_c(exact)= \ln(1+\sqrt(8))=1.342454 \ldots$}}
\vspace{3ex}
\hspace*{-1cm}\begin{tabular}{r|lrll|lrll|lrll}
\multicolumn{1}{c|}{range L's}  &
\multicolumn{4}{c|}{$\beta^C_{\rm max}(L)=\beta_c+a_1/L+a_2/L^2$} &
\multicolumn{4}{c|}{$\beta^{\chi}_{\rm max}(L)=\beta_c+a_1/L+a_2/L^2$} &
\multicolumn{4}{c}{$\beta^B_{\rm max}(L)=\beta_c+a_1/L+a_2/L^2$}  \\ \hline
& \multicolumn{1}{c}{Q} & \multicolumn{1}{c}{$\beta_c$} & \multicolumn{1}{c}{$a_1$}
& \multicolumn{1}{c|}{$a_2$} &
\multicolumn{1}{c}{Q}&\multicolumn{1}{c}{$\beta_c$}&\multicolumn{1}{c}{$a_1$}
&\multicolumn{1}{c|}{$a_2$}&
\multicolumn{1}{c}{Q}&\multicolumn{1}{c}{$\beta_c$}&\multicolumn{1}{c}{$a_1$}
&\multicolumn{1}{c}{$a_2$} \\ \hline \hline
              &  &  &  &  &  &  &   &  & & & &  \\
  84 -- 350 & 0.11 & 1.342494(38) & -0.219(15) & -25.5(1.1)& 0.13 & 1.342478(38)
   & -0.208(15)& -27.2(1.1)& 0.13 & 1.342481(38) & -0.210(15)& -28.3(1.1) \\
              &  &  &  &  &  &  &   &  &  \\
 100 -- 350 & 0.72  & 1.342423(46) & -0.187(19)& -28.6(1.6)& 0.73 & 1.342408(46)
   & -0.177(18)& -30.3(1.6) & 0.73 & 1.342412(46) & -0.180(18)& -31.3(1.6)\\
            &  &  &  &  &  &  &   &  & & & &  \\
\end{tabular}
\label{tab:beta-fits}
\end{table}

\subsection{Analysis of $C_{\rm max}$, $\chi_{\rm max}$ and
            $B_{\rm min}$}

The results of the fits to the specific heat and susceptibility maxima,
$C_{\rm max}$ and $\chi_{\rm max}$, together with the
kurtosis minimum are summarized in Table~\ref{tab:extrema_fits}.
As before, we show the fits for two ranges of lattice sizes.
Notice that the linear correction coefficients, $c_1$ and $e_1$,
are two orders of magnitude larger than the coefficients, $c_2$ and $e_2$,
of the dominant contribution $L^2$. This makes necessary to adjust the data
to the ansatz $C_{\rm max}(L) = c_0 + c_1 \, L + c_2 \, L^2$, and allows
to estimate the corrections to the leading term.

In simulations with PBC, the correction to the leading term is
of the order $\gamma_1/L^2$. If we fit our specific heat data in the range
$L=126-350$ to the ansatz $C_{\rm max}(L) = \gamma_0 + \gamma_1 / L^2 + \gamma_2 \,
L^2$, the goodness-of-fit is $Q=0.0003$ with an absurdly high value
for $\gamma_1$. On the other hand, if we do not allow for a correction
term and fit the data to $C_{\rm max}(L) = \gamma_0 + \gamma_2 \, L^2$, the
goodness-of-fit turns out to be 0.

The work of Medved~\cite{M} shows that the coefficient of $L^2$ in
the finite size scaling of $C_{\rm max}$ is related to the latent
heat, $\Lambda_{\rm FBC}$, via $c_2=(\Lambda_{\rm FBC} \, \beta_c /
2)^2$. In fact, it is the same relationship that holds for periodic boundary
conditions~\cite{Challa,BKS,Janke}. If we use our estimation $c_2=0.0389(29)$
from Table~\ref{tab:extrema_fits} and $\beta_c=\ln(1+\sqrt(8))$,
we obtain for the latent heat
\begin{equation}
  \Lambda_{\rm FBC} = 0.294(11).
\end{equation}

Another way of estimating the latent heat is from the direct calculation,
right at the transition, of the internal energies per site
of the ordered and disordered phases, $e_{\rm ord}=E_{\rm ord}/V$, and
$e_{\rm dis}=E_{\rm dis}/V$.
Of course, the latent heat is just $\Lambda=e_{\rm dis} \, - \, e_{\rm ord}$. Lee and
Kosterlitz proposed~\cite{Kosterlitz} to reweight a given
energy histogram until both peaks have equal height. The locations
of the two maxima in the histogram can be taken as finite size estimates,
$e_{\rm o}(L)$ and $e_{\rm d}(L)$, for the infinite-volume limits at $\beta_c$ of
$e_{\rm ord}$ and $e_{\rm dis}$. The scaling of $e_{\rm o}(L)$ and $e_{\rm d}(L)$ for
fixed boundary conditions~\cite{M} as well as periodic boundary
conditions~\cite{Kosterlitz} is $e_{\rm o}(L)=e_{\rm ord}+O(1/L)$
and $e_{\rm d}(L)=e_{\rm dis}+O(1/L)$.

We smoothed~\cite{NR} our energy histograms to reduce the noise
and searched for $e_{\rm o}(L)$ and $e_{\rm d}(L)$. Table~\ref{tab:e-ord-dis} shows
the estimations that we found. Fitting them to the ansatz
$e_{\rm o}(L)=e_{\rm ord}+k_1/L$ and $e_{\rm d}(L)=e_{\rm dis}+k_2/L$,
we obtained $e_{\rm ord}=-1.6032(48)$ and $e_{\rm dis}=-1.3114(92)$,
with goodness-of-fit~\cite{NR} $Q=1$ and $Q=0.9$ respectively. Consequently
another estimation for the latent heat is
\begin{equation}
  \Lambda = 0.292(10).
\end{equation}
The agreement with our previous estimation could not be better: it is quite comforting.

R.J. Baxter~\cite{Baxter1,Baxter2} derived an analytical expression for the
latent heat of the $q$-state Potts model assuming periodic boundary conditions.
Numerical evaluations of his expression are tabulated in
Wu~\cite{Wu} and Janke~\cite{Janke}. For $q=8$, the latent heat for
the Potts model with periodic boundary conditions is $\Lambda_{\rm PBC} = 0.486358\ldots$
Obviously our estimations of the latent heat do not coincide with
this value, but it should not be so surprising in view of Fig.~\ref{fig:hist2}, where
it can be seen that, for $L=100$, the distance between peaks for P.B.C. is
so different from the distance between peaks for F.B.C. Although such differences could
tend towards the same value with $L \rightarrow \infty$, our analysis indicates that in fact
they do not.

Notice that, unlike the latent heat, the analytically known infinite-volume critical
inverse temperature, $\beta_c = \ln(1+\sqrt(q))$, for the $q$-state Potts model is
derived~\cite{Potts,Wu2,Hintermann,Baxter2} using the self-duality property of
the model, which is independent of boundary conditions when $L \rightarrow
\infty$. Let us recall that our estimations of $\beta_c$ are consistent
with $\beta_c=1.342454\ldots$

%
%
\begin{table}[htbp]
\centering \caption[{\em Nothing.}]
 {\em  Fits on the extrema of $C_{\rm max}$, $\chi_{\rm max}$ and
       $B_{\rm min}$.}
\vspace{3ex}
\hspace*{-1cm}\begin{tabular}{r|llll|llll|llll}
\multicolumn{1}{c|}{range L's} &
\multicolumn{4}{c|}{$C_{\rm max}(L)    = c_0 + c_1 \, L + c_2 \, L^2$}  &
\multicolumn{4}{c|}{$\chi_{\rm max}(L) = e_0 + e_1 \, L + e_2 \, L^2 $} &
\multicolumn{4}{c} {$B_{\rm min}(L) =  B_0 + B_1 / L + B_2 / L^2$} \\ \hline
& \multicolumn{1}{c}{Q} & \multicolumn{1}{c}{$c_0$} & \multicolumn{1}{c}{$c_1$}&
  \multicolumn{1}{c|}{$c_2$} &
 \multicolumn{1}{c}{Q} & \multicolumn{1}{c}{$e_0$} & \multicolumn{1}{c}{$e_1$}&
  \multicolumn{1}{c|}{$e_2$} &
 \multicolumn{1}{c}{Q} & \multicolumn{1}{c}{$B_0$} & \multicolumn{1}{c}{$B_1$}&
  \multicolumn{1}{c}{$B_2$} \\ \hline \hline
 &  &  &  &  &  &  &  & & & & & \\
  100 -- 350  & 0.012 & 254(25) & -4.24(35) & 0.0342(11) &  & & &
              & 0.011 & 0.65461(40) & 1.52(13) & -92.1(9.2) \\
 &  &  &  &  &  &  &  & & & & & \\
  126 -- 350  & 0.15  & 427(65) & -6.14(75) & 0.0389(29) & 0.033 & 766(101) &
               -11.9(1.2) & 0.0691(31) & 0.11 & 0.65295(72) &
               2.19(28) & -153(24) \\
 &  &  &  &  &  &  &  & & & & & \\
  150 -- 350  &   &  &  &  & 0.38 & 1262(203) &
               -16.7(2.1) & 0.0798(49) &  &  &  &
  \\
\end{tabular}
\label{tab:extrema_fits}
\end{table}

%

%
%
\begin{table}[htbp]
\centering \caption[{\em Nothing.}]
 {{\em Finite size estimates $e_{\rm o}(L)$ and $e_{\rm d}(L)$.
 They are obtained by reweighting the energy histograms until both peaks
 have equal heights. The infinite-volume ordered and disordered
 energies are estimated from the ansatz $e_{\rm o}(L)=e_{\rm ord}+k_1/L$
 and $e_{\rm d}(L)=e_{\rm dis}+k_2/L$.}}
\vspace{3ex}
\begin{tabular}{c|ll}
\multicolumn{1}{c|}{$L$} &
\multicolumn{1}{c}{$e_{\rm o}$} &
\multicolumn{1}{c}{$e_{\rm d}$} \\
\hline
100 &  -1.580(11)  & -1.4167(74) \\
126 &  -1.586(10)  & -1.398(20)  \\
150 &  -1.587(18)  & -1.398(13)  \\
226 &  -1.5965(93) & -1.3623(91) \\
250 &  -1.5944(83) & -1.362(14)  \\
278 &  -1.5970(39) & -1.350(17)  \\
300 &  -1.5960(27) & -1.3452(98) \\
350 &  -1.5958(31) & -1.337(13) \\
\multicolumn{1}{c|}{$\vdots$} & \multicolumn{1}{c}{$\vdots$} &
\multicolumn{1}{c}{$\vdots$}  \\
$\infty$ & -1.6032(48) & -1.3114(92) \\
\end{tabular}
\label{tab:e-ord-dis}
\end{table}

\section{Conclusions}

The first order phase transition finite-size scaling laws for
Fixed Boundary Condition lattices of Borgs-Kotecky-Medved have been
presented, tested and shown to be the only ones that hold for the
2d 8-state Potts model.

It is clear from our analysis that Monte Carlo simulations for FBC
are necessarily going to be much more time consuming than those
for PBC, since for PBC the system sets into the finite-size
scaling region as $\beta_c(L)=\beta_c(\infty)+\theta_1/L^d$, while for
FBC it does it at the slower pace of
$\beta_c(L)=\beta_c(\infty)+a_1/L$. Besides, we have found that
the latent heat is affected by the boundaries.

It is a pleasure to thank W. Janke and A. Salas  for ve\-ry
sti\-mu\-la\-ting dis\-cus\-sions.  Financial support from CICYT
contracts AEN98-0431, AEN99-0766, is acknowledged.

\end{document}